\documentclass[10pt,letterpaper]{article}
\usepackage{opex3}
\usepackage{array}
\usepackage{subfig}
\usepackage{bm}
\usepackage{amsfonts}
\usepackage{amsmath}
\usepackage{epsfig}
\def\SX{{{\cal S}_X}}

\def\bx{{\bar x}}

\def\bE{\mathbb{E}}

\def\be{\begin{equation}}
\def\ee{\end{equation}}
\def\ba{\begin{align}}
\def\defeq{\buildrel \rm def \over =}
\newcommand{\argmax}[1]{\underset{#1}{\operatorname{arg}\operatorname{max}}\;}
\newcommand{\argmin}[1]{\underset{#1}{\operatorname{arg}\operatorname{min}}\;}
\newcommand{\norm}[1]{\left\lVert#1\right\rVert_2}

\def\MPE{P_e^{(min)}}
\def\cR{{\cal R}}
\def\cRm{{{\cal R}_m}}
\def\cRmp{{{\cal R}_{m^\prime}}}
\def\SX{{{\cal S}_X}}
\def\tm{{\tilde m}}
\begin{document}
\title{\bf Asymptotics of Bayesian Error Probability and Rotating-PSF-Based Source Super-Localization in Three Dimensions}
\author{S.~Prasad}%
\address{Department of Physics
and Astronomy, University of New Mexico, Albuquerque, New Mexico 87131}
\email{sprasad@unm.edu}

\begin{abstract}
We present an asymptotic analysis of the minimum probability of error (MPE)
in inferring the correct hypothesis in a Bayesian multi-hypothesis testing (MHT) formalism
using many pixels of data that are corrupted by signal dependent shot noise, sensor read noise, and 
background illumination. We perform this error analysis for a variety
of combined noise and background statistics, including a pseudo-Gaussian distribution that
can be employed to treat approximately the photon-counting statistics of signal and background as well as purely Gaussian
sensor read-out noise and more general, exponentially peaked distributions.
We subsequently apply the MPE asymptotics to characterize the minimum conditions needed to
localize a point source in three dimensions by means of a rotating-PSF imager and compare its performance
with that of a conventional imager in the presence of background and sensor-noise fluctuations.
In a separate paper \cite{SP14}, we apply the formalism to the related but qualitatively different problem
of 2D super-resolution imaging of a closely spaced pair of point sources in the plane of best focus.
\end{abstract}


\section{Introduction}
Spatial localization of single molecules well beyond the diffraction limit is an important task of many flourescence-based
bioimaging techniques that are presently available \cite{HBZ09,PDML10}. This includes both full 3D localization as well as motion tracking.
The unprocessed image data typically provide initial information about the coarser ranges to which the position coordinates
of a single molecule in question are confined, but a more careful post-processing of the image data involving either
a point-spread function (PSF) fitting \cite{Bobroff86} or a statistical approach \cite{CWG01} or
a combined approach using Bayesian inference \cite{YBFK08} and mean-squared displacement analysis \cite{MGM12,CRM12}
can reveal the location of and track \cite{GM14} one or more fluorescing molecules with much finer precision.
It is possible by these methods to achieve a precision 10-20 times smaller - occasionally even 100 times smaller \cite{YFM03}
- than the standard Abbe diffraction limit, $\delta x \geq 0.61\lambda/NA$,
of a microscope operating at an observing wavelength $\lambda$ and with numerical aperture $NA$.

The typical error analysis for biomolecular localization has considered the mean-squared error (MSE)
in estimating the location, either via a centroid-based PSF model fitting \cite{TLW02} or a
Cramer-Rao-bound (CRB) based minimum-estimator-variance analysis \cite{ORW04,RWO06, RWO05}.
These analyses are local and incomplete, however, since they fail to account for any prior knowledge
one may have about the location based on low-resolution image data, which may be critical
to the localization accuracy at low signal strengths. A
local analysis also excludes, by its very definition, any higher-order sensitivity criteria
that would permit a more accurate treatment of certain problems for which a first-order sensitivity
metric like the Fisher information (FI), on which the CRB is based \cite{Kay93}, vanishes identically. An important
problem of this kind is the quantification of the error in the axial localization of a molecule at zero defocus
using any defocus-symmetric PSF, such as the conventional Airy-disk PSF \cite{RWO05}.
It seems to us more sensible, although
theoretically less tractable, to pose the problem of super-localizing single molecules to sub-diffractive
precision in a Bayesian framework where any prior knowledge may be incorporated in a statistical
formalism in a non-local manner.

In the analysis presented here, the acquired image data automatically satisfy a Bayesian protocol in
which the coarse spatial ranges bounding the source coordinates serve
to provide an initial, or prior, statistical distribution of these coordinates, one that without additional
spatial information can be regarded as being
uniformly random over these ranges. Under a sufficiently high signal-to-noise ratio (SNR)
of fluorescence detection, the data $X$ contain, via their specific statistical noise model, detailed information about the source location.
Their conditional statistics, specified by a probability density function (PDF), $P(x\mid m)$,
conditioned on the knowledge of a specific source location $m$, can be exploited to improve upon the location uncertainty to
a degree that depends on directly on the source flux and background levels.

The problem of localization with sub-diffractive errors amounts, in our multi-hypothesis-testing (MHT) based Bayesian view \cite{vTB13},
to determining the source location by means of the posterior distribution, $P(m|X)$, with a greater precision
than that contained in the uniform spatial distribution of the prior $p_m$, $m=1,\ldots,M$.
Indeed, the mean and mode of the posterior provide two different excellent estimates of source location. The mean-squared error
(MSE) of the first of these estimators, the so-called minimum MSE (MMSE), upper-bounds the best achievable precision theoretically
possible with any image processing protocol. The maximum {\it a posteriori} (MAP) estimator provides a second
fundamental metric of minimum error, namely the minimum probability of error (MPE) in estimating
the correct source position from among a set of $M$ {\it a priori} random possible positions,
\be
\label{e0}
\MPE=1-\bE\left[P(\hat m_{MAP}|X)\right],
\ee
where $\hat m$ is the MAP estimator,
\be
\label{e01}
\hat m_{MAP}=\argmax{m=1,\ldots,M}P(m|X).
\ee
While the two metrics are in general not related to each other,
the MMSE is lower-bounded by the so-called Ziv-Zakai bound \cite{ZZ69} which is expressed in
terms of the MPE of a related binary-hypothesis testing problem.
The MPE metric also provides a useful relationship between Bayesian inference
and statistical information via the Fano bound and its generalizations \cite{CT91}.
The present work extends our earlier study \cite{SP12} of localization accuracy based
on both the MPE and MMSE metrics, which were shown to be related closely for
a highly sensitive Bayesian detector, from a few sensor pixels to the asymptotic domain of many
sensor pixels. It differs, however, from the more standard asymptotic analyses of MHT \cite{LJ97}
in which one assumes that many statistically identical data frames are present,
by addressing the experimentally more realistic context of many image pixels that sample a spatially varying
PSF, typically with a single peak so the pixels progressively farther away from the peak contain
progressively less signal.

The work reported here is a comprehensive analysis of the MPE within the MHT protocol,
as applicable to problems involving imaging in the presence of photon and sensor noise.
We have also developed approximate treatments of the MMSE metric, but the use
of that metric to characterize the error of point-source super-localization
will be presented elsewhere. We focus here exclusively on an MPE-based Bayesian error analysis
of this problem.

In Sec. 2, we present a brief review of the general MHT error analysis.
In Sec. 3, we discuss the problem of localization of a single point source
from the MHT perspective, focusing on the calculation of the MPE in a MAP-based
detection protocol for localization under purely Gaussian additive sensor noise statistics.
For such noise statistics, it is possible to perform a simple asymptotic analysis of the MPE
of localization and show that the MPE may be expressed as a sum of complementary error functions,
in which at high SNR only one term tends to dominate all others that can thus be dropped.
Section 4 generalizes the calculations of the previous section to more general noise statistics that include
the fluctuations of a spatially uniform mean background
and photon noise from the signal itself, both described by Poisson statistics.
Under a simplifying approximation that all three noise statistics can be combined at each image pixel into
a single pseudo-Gaussian statistics with the mean being equal to the sum of the (spatially varying) signal and
(spatially uniform) background mean values, and variance being equal to the sum of
the sensor noise variance and the signal and background mean values. Under Poisson statistics, since the two mean values
also furnish the corresponding variances, this simplifying approach
may be justified under conditions of either a large signal mean value or
a large background mean value per pixel.
A continuity correction \cite{Roussas97}, not implemented here, can improve the pseudo-Gaussian approximation still further
for the combined noise statistics.
In Sec.~5, these considerations are generalized further to include any exponentially peaked statistical distributions,
for which asymptotically valid approximations
of the kind covered in Sec.~4 are once again possible.
The next section presents in graphical form our numerically obtained exact and asymptotic results
for the MPE for the problem of super-localization of single point sources by two different
classes of imaging microscopes, one based on conventional clear-aperture imaging and the other
on a recently proposed \cite{SP13} pupil-phase-engineered PSF that encodes defocus via its rotation in the image
plane and is thus naturally suited to perform axial localization.
Detailed comparisons of the performances of the two imagers with respect to the problem of
full 3D localization of point sources are presented and discussed in this section.
Some concluding remarks appear in Sec.~7.

\section{Minimum Probability of Error for M-ary Hypothesis Testing}\label{sec:intro}

For Bayesian discrimination among $M$ different hypotheses, labeled by the index $m$, which takes
values $m=1,\ldots,M$, expression (\ref{e0}) for
the MPE in inferring the correct hypothesis from data $X$, drawn from the
set $\SX$, may be reformulated by means of the Bayes rule as
\be
\label{e1}
\MPE=1-\sum_{m=1}^M p_m\int_\cRm dx \, P(x\mid m),
\ee
where $\cRm$ is the decision region in the data set $\SX$ for the $m$th hypothesis. The MAP
criterion, namely
\be
\label{e1a}
\cRm=\left\{x\, \mid\, P(x\mid m)\, p_m \geq P(x\mid m^\prime)\,p_{m^\prime}, \ \forall m^\prime\neq m\right\},
\ee
defines the decision regions for which the error probability (\ref{e1}) takes its minimum value.
The $M$ different hypotheses exhaust all possible decisions from any data outcome, {\it i.e.,}
$\SX =\cup_{m=1}^M \cRm$. Since $\int_\SX dx\, P(x\mid m) = 1$ for any $m$, we may express Eq.~(\ref{e1})
more conveniently as
\be
\label{e2}
\MPE=\sum_{m=1}^M p_m\sum_{m^\prime\neq m}\int_\cRmp dx \, P(x\mid m).
\ee

We shall assume that the data space is real and multi-dimensional, as the image data are typically collected
at a number of pixels, say $N$, in the sensor plane. Thus, $\SX \subset \mathbb{R}^{N}$.
For $N>>1$, as is typically the case, $\MPE$ may be evaluated approximately via an asymptotic analysis. In fact, one
may show quite easily that for a fixed value of $m$, the sum over $m^\prime$ in Eq.~(\ref{e2}) may be replaced in this case by a single
term $\tilde m$, which labels the decision region that is the ``closest" to $\cRm$ in the following sense:
\be
\label{e3}
\tilde m = \argmax{m^\prime\neq m} \max_{x\in \cRmp}\left\{P(x\mid m)\right\}.
\ee
The MPE is thus accurately approximated by the asymptotic expression
\be
\label{e2a}
\MPE=\sum_{m=1}^M p_m\int_{{\cal R}_{\tilde m}} dx \, P(x\mid m).
\ee

\section{Gaussian Conditional Data Statistics}

For Gaussian conditional data statistics, namely
\be
\label{e4}
P(x\mid m)={1\over(2\pi\sigma^2)^{N/2}}\exp\left[-(1/2)\norm{x-x_m}^2/\sigma^2\right],
\ee
where $x_m$ denotes the mean value of the data vector under hypothesis $m$ and $\sigma^2$ the variance
of data at each pixel, the definition of $\tilde m$ may equivalently be stated as
\be
\label{e5}
\tilde m = \argmin{m^\prime\neq m} \min\left\{E_m(x)\vert x\in \cRmp\right\},
\ee
where up to an additive constant $E_m(x)$ is simply proportional to $-\ln [P(x|m)\, p_m]$,
\be
\label{e6}
E_m(x)\defeq \norm{x-x_m}^2-2\sigma^2\ln p_m.
\ee

For a given $m$, we determine $\tilde m$ by first mapping out the boundary between the decision region $\cRm$
and other decision regions and then finding that decision region for which $E_m(x)$ takes the smallest
possible value at the boundary. This is an optimization problem that is easily solved by requiring that at
the boundary between $\cRm$ and $\cRmp$, $E_m(x)=E_{m^\prime}(x)$, {\it i.e.},
\be
\label{e7}
\norm{x-x_m}^2 = \norm{x-x_m-\delta x_{mm^\prime}}^2+2\sigma^2\ln (p_m/p_{m^\prime}),
\ee
where $\delta x_{mm^\prime}$ simply denotes the separation vector between the mean data values for the $m$th and $m^\prime$th hypotheses,
\be
\label{e8}
\delta x_{m^\prime m}\defeq x_{m^\prime}-x_m.
\ee
Expanding the squared norm on the right hand side (RHS) of Eq.~(\ref{e7}) using the identity $\norm{a-b}^2=\norm{a}^2+\norm{b}^2
-2a^Tb$, where $a$ and $b$ are two column vectors, we see that the boundary between $\cRm$ and $\cRmp$
is described by the equation
\be
\label{e9}
\left(x-x_m-\gamma{\delta x_{m^\prime m}\over 2}\right)^T \delta x_{m^\prime m} =0,
\ee
where $\gamma$ is defined by the relation
\be
\label{e9a}
\gamma=1+{2\sigma^2\ln (p_m/p_{m^\prime})\over\norm{\delta x_{mm^\prime}}^2}.
\ee
Equation (\ref{e9}) defines a hyperplane that passes through the point $x_m+\gamma\delta x_{mm^\prime}/2$ on
the mean-to-mean separation vector $\delta x_{m^\prime m}$ and is orthogonal
to that vector. Clearly, over this hyperplane the common value of $E_m$ and $E_{m^\prime}$ has its minimum
at this point, namely at $x=x_m+\gamma\delta x_{m^\prime m}/2$, given by
$\min_{x\in \cRmp}(E_m)=F_{mm^\prime}^2-2\sigma^2\ln p_m$, where
\be
\label{e10}
F_{mm^\prime}^2\defeq {\gamma^2\over 4}\norm{x_m-x_{m^\prime}}^2.
\ee
The index of the ``closest" decision region
to $\cRm$, as defined by Eq.~(\ref{e5}), is then the argument of the minimum value of $\min_{x\in\cRmp}(E_m)$ over
all $m^\prime\neq m$, {\it i.e.},
\be
\label{e11}
\tilde m = \argmin{m^\prime\neq m} \left(F_{mm^\prime}^2-2\sigma^2\ln p_m\right).
\ee

In view of the PDF (\ref{e7}), the asymptotically correct expression (\ref{e2a}) may be evaluated
approximately by transforming the integral on its RHS, for each $m$ value, to a coordinate system in the $N$-dimensional data space for which one
of the coordinate unit vectors, say $\hat t$, is chosen to be along the separation vector $x_{\tilde m}-x_m$ and the remaining $(N-1)$
coordinate axes are chosen to span the hyperplane, orthogonal to $\hat t$, that separates the decision region $\cRm$ from its closest
neighbor ${\cal R}_{\tilde m}$. The deviation of a data vector from its mean value, $(x-x_m),$ may then be expressed
in the new coordinate basis as $x-x_m=x_t \hat t +x_\perp$, where $x_\perp$ is the
projection of $x$ in the $(N-1)$ dimensional hyperplane orthogonal to $\hat t$, {\it i.e.,} $x_\perp^T \hat t=0$.
This transformation allows us to express the squared norm (\ref{e6}) as
\be
\label{e12}
E_m(x)=x_t^2 +\norm{x_\perp}^2,
\ee
and the PDF (\ref{e4}) as
\be
\label{e13}
P(x\mid m)={1\over(2\pi\sigma^2)^{N/2}}\exp\left[-(1/2)\left(x_t^2+\norm{x_\perp}^2\right)/\sigma^2\right].
\ee

On substitution of this form for $P(x\mid m)$ inside the integral in the asymptotic expression (\ref{e2a}) and integrating the
variable $x_t$ from $F_{m\tilde m}$ to $\infty$ and the orthogonal projection $x_\perp$
over the full hyperplane containing it, we have the simple result
\be
\label{e14}
\MPE=\sum_{m=1}^M p_m{1\over (2\pi\sigma^2)^{1/2}}\int_{F_{m\tilde m}}^\infty \exp[-(1/2)x_t^2/\sigma^2],
\ee
in which we have used the simple integral identity,
$$\int_{-\infty}^\infty dx \, \exp[-(1/2)x^2/\sigma^2] = (2\pi\sigma^2)^{1/2},$$
$(N-1)$ times to integrate over the $(N-1)$ nonzero, mutually orthogonal components of the vector $x_\perp$.
The remaining integral in expression (\ref{e14}) may be evaluated in terms of the complementary error function as
\be
\label{e15}
\MPE={1\over 2}\sum_{m=1}^M p_m{\rm erfc\,}(F_{m\tilde m}/\sqrt{2\sigma^2}),
\ee
with ${\rm erfc}$ defined as
\be
\label{e16}
{\rm erfc\,}(u)\defeq {2\over\sqrt{\pi}}\int_u^\infty \exp(-x^2)\, dx.
\ee

For sensitive detectors, the quantity, $F_{m\tilde m}/\sqrt{2\sigma^2}$, is likely to be large, as it is
proportional to the SNR for any $m$. In that case, expression (\ref{e15})
simplifies still further since the asymptotically valid approximation for erfc may then be used,
\ba
\label{e17}
\MPE=&{1\over \sqrt{2\pi}}\sum_{m=1}^M p_m{\sigma\over F_{m\tilde m}}\exp[-(1/2)F_{m\tilde m}^2/\sigma^2]\nonumber\\
    =&{1\over \sqrt{2\pi}}\sum_{m=1}^M p_m{2\sigma\over \gamma\norm{x_{\tilde m}-x_m}}\exp\left[-\gamma^2{\norm{x_{\tilde m}-x_m}^2
\over 8\sigma^2}\right].
\end{align}

\section{Pseudo-Gaussian Conditional Data Statistics}

A similar but considerably more involved treatment of the MPE may be given for a pseudo-Gaussian conditional data PDF that
accurately describes the statistics of image data acquired under combined photon-number fluctuations and sensor read-out noise,
at least at large photon numbers. Let the data $x$, given hypothesis $m$, be distributed according to the PDF
\be
\label{e18}
P(x\mid m) ={1\over (2\pi)^{N/2}{\rm det\,}^{1/2}(\Sigma_m)}\exp[-(1/2)(x^T-x_m^T)\Sigma_m^{-1}(x-x_m)].
\ee
where, under the condition of statistically independent data pixels, the data covariance matrix is a diagonal matrix
of the form\footnote{
We use here a shorthand notation, diag$(v)$, for specifying a diagonal matrix whose
diagonal elements are the elements of $v$ taken in order. We shall also use the notation, diag$(u/v)$, to denote
the diagonal matrix of elements that are ratios of the corresponding elements of the vectors $u$ and $v$. In Matlab, this would
be the element-wise quotient, $u./v$, of the two vectors of which the diagonal matrix is formed.}
\be
\label{e19}
\Sigma_m ={\rm diag}(\sigma^2+x_m),
\ee
where $\sigma_2$ and $x_m$ denote, as before, the variance of sensor read-out noise
and the mean data vector, respectively, given the hypothesis $m$.
The variance of the pseudo-Gaussian PDF (\ref{e19}) is the sum of the Gaussian
read-out noise variance and the variance of the shot noise corresponding to
Poisson photon-number fluctuations, the latter being equal to the mean photon
number at any pixel.

Under asymptotic conditions, as for Gaussian data statistics, the most significant
contributions to the MPE from the $m$th term in the sum (\ref{e2a}) come from the
vicinity of the point, $x_*$, on the boundary between $\cRm$ and $\cR_{\tilde m}$
where $P(x\mid m)$ has its largest value. This point does not, however, lie
on the line joining the centers of the two decision regions nor is the boundary between
two decision regions a hyperplane in general. Rather one
must perform a constrained maximization of $P(x\mid m)$, or equivalently a
minimization of $-\ln P(x\mid m)$,
\be
\label{e20}
-\ln P(x\mid m)=(1/2)\Big\{(x^T-x_m^T)\Sigma_m^{-1}(x-x_m)+\ln [(2\pi)^N{\rm det}\,\Sigma_m]\Big\},
\ee
subject to the constraint that $x$ be
on the boundary, {\it i.e.}, $-\ln [P(x\mid m)\, p_m]=-\ln [P(x\mid \tilde m)\, p_\tm]$, according to the
MAP decision rule underlying the MPE expression. In view of the form (\ref{e20})
for the negative log-likelihood function (LLF), this amounts, via the use of a
Lagrange multiplier $\lambda$, to the minimization,
\ba
\label{e21}
\min_{x} (x^T-x_m^T)\Sigma_m^{-1}(x-x_m)& -\lambda \big[(x^T-x_m^T)\Sigma_m^{-1}(x-x_m)\nonumber\\
                 &-(x^T-x_\tm^T)\Sigma_\tm^{-1}(x-x_\tm)\big],
\end{align}
from which, for brevity, we have dropped certain logarithmic terms that do not depend on $x$.

The minimum of the quadratic form (\ref{e21}) is easily determined by taking its gradient wrt $x$
and setting it to zero at $x=x_*$,
\be
\label{e22}
(1-\lambda)\Sigma_m^{-1}(x_*-x_m)+\lambda \Sigma_\tm^{-1} (x_*-x_\tm)=0.
\ee
This equation is readily solved as a simple matrix equation for $(x_*-x_m)$ by writing $(x_*-x_\tm)=(x_*-x_m)-\delta x_{\tm m}$, where
$\delta x_{\tm m}$ is defined by relation (\ref{e8}) and then
combining the two terms containing it. The solution may be simplified by employing the definition (\ref{e19}) for
the diagonal covariance matrices $\Sigma_m$ and $\Sigma_\tm$ and then performing simple algebra,
\be
\label{e23}
(x_*-x_m)={\rm diag}\left[1+{(1-\lambda)\over\lambda}{(\sigma^2+x_\tm)\over (\sigma^2+x_m)}\right]^{-1}\delta x_{\tm m}.
\ee
A similar expression for $(x_*-x_\tm)$ also follows from (\ref{e23}),
\ba
\label{e24}
(x_*-x_\tm)&= (x_*-x_m)-\delta x_{\tm m}\nonumber\\
&=-{\rm diag}\left[1+{\lambda\over(1-\lambda)}{(\sigma^2+x_m)\over (\sigma^2+x_\tm)}\right]^{-1}\delta x_{\tm m}.
\end{align}
These expressions, when plugged into the constraint, $-\ln [P(x\mid m)\, p_m]=-\ln [P(x\mid \tm)\, p_\tm]$, with the negative LLF given by
expression (\ref{e20}), yield an equation for $\lambda$, which may be simplified to the form
\ba
\label{e25}
\delta x_{\tm m}^T{\rm diag}\,&\left\{ {\lambda^2(\sigma^2+x_m)-(1-\lambda)^2(\sigma^2+x_\tm)\over
[\lambda (\sigma^2+x_m)+(1-\lambda)(\sigma^2+x_\tm)]^2}\right\}\delta x_{\tm m}\nonumber\\
&=\ln {p_m^2{\rm det}\, \Sigma_\tm\over p_\tm^2{\rm det}\, \Sigma_m}.
\end{align}

Equation (\ref{e25}) is sufficiently complicated that its solution, $\lambda$, cannot be evaluated in
closed form. However, under asymptotic conditions of many pixels, $N>>1$, and large source flux, $\norm{x_m}>> N\sigma^2>>1$,
for which we expect the inequalities, $1<<\norm{\delta x_{\tm m}} <<\norm{\sigma^2+x_m}$, to hold,
we may write $(\sigma^2+x_\tm)=(\sigma^2+x_m)+\delta x_{\tm m}$ and expand the element-wise quotient, $r_m$, of the two
vectors inside the curly braces in Eq.~(\ref{e25}) in powers of $\delta x_{\tm m}/(\sigma^2+x_m)$
to obtain
\ba
\label{e26}
r_m&\defeq
 {\lambda^2(\sigma^2+x_m)-(1-\lambda)^2(\sigma^2+x_\tm)\over
[\lambda (\sigma^2+x_m)+(1-\lambda)(\sigma^2+x_\tm)]^2}\nonumber\\
&={(2\lambda-1)(\sigma^2+x_m)-(1-\lambda)^2\delta x_{\tm m}\over
[\lambda (\sigma^2+x_m)+(1-\lambda)(\sigma^2+x_\tm)]^2}\nonumber\\
&= {(2\lambda-1)\over (\sigma^2+x_m)}\left[1-{(1-\lambda)^2\over (2\lambda-1)}
{\delta x_{\tm m}\over (\sigma^2+x_m)}\right]\nonumber\\
&\quad\times \left[1-2(1-\lambda)
{\delta x_{\tm m}\over (\sigma^2+x_m)}+O(\delta^2)\right]\nonumber\\
&={(2\lambda-1)\over (\sigma^2+x_m)}\left[1-{(3\lambda-1)(1-\lambda)\over (2\lambda-1)}
{\delta x_{\tm m}\over (\sigma^2+x_m)}+O(\delta^2)\right],
\end{align}
where $\delta$ is a shorthand notation for the order of the asymptotically small elements of the element-wise
quotient vector $\delta x_{\tm m}/(\sigma^2+x_m)$.
The RHS of Eq.~(\ref{e25}) may be similarly expanded by writing the logarithm of the ratio of
the determinants of the
diagonal covariance matrices, $\Sigma_\tm$ and $\Sigma_m$, as a sum of the logarithms of the ratios of
their diagonal elements, ratios that, in the asymptotic limit, differ from 1 by the small ratios of the corresponding elements
of $\delta x_{\tm m}$ and $(\sigma^2+x_m)$.
On performing this expansion to the lowest order in these ratios
and substituting expression (\ref{e26}) in Eq.~(\ref{e25}), we obtain a simple equation for $\lambda$ in
this order,
\be
\label{e27}
(2\lambda-1)\delta x_{\tm m}^T{\rm diag}\, (\sigma^2+x_m)^{-1}\delta x_{\tm m}=\sum_{i=1}^N
{\delta x_{\tm mi}\over (\sigma^2+x_{mi})}+2\ln {p_m\over p_\tm},
\ee
from which $\lambda$ may be determined as
\be
\label{e28}
\lambda={1\over 2}+{1\over 2}{\sum_{i=1}^N{\delta x_{\tm mi}\over(\sigma^2+x_{mi})}+2\ln{p_m\over p_\tm}\over
\sum_{i=1}^N{\delta x^2_{\tm mi}\over(\sigma^2+x_{mi})}}.
\ee

The contribution of the $m$th term in the sum (\ref{e2a}) for the MPE may now be calculated
in the asymptotic regime by recognizing that the integral of $P(x\mid m)$ over $\cR_\tm$
is dominated by its maximum value, $P(x_*\mid m)$, on the boundary between $\cRm$ and
$\cR_\tm$. The calculation can be made more precise by first determining the unit normal $n_*$ to the boundary at
the point $x_*$, and then resolving the vector,
\be
\label{e29}
v\defeq \Sigma_m^{-1/2}(x-x_m),
\ee
whose negative squared norm occurs in the exponent of the Gaussian form of $P(x\mid m)$,
along $n_*$ and orthogonal to it.
The integral of $P(x\mid m)$ over $\cR_\tm$ may then be performed by fixing the origin
of an orthonormal coordinate system at $x_*$, with coordinate axes that are along $n_*$ and orthogonal
to it.
The $N$-dimensional integral in expression (\ref{e2a}) then reduces, approximately,
to the product of a Gaussian integral along $n_*$ from 0 to $\infty$ and
the remaining $(N-1)$ Gaussian integrals along mutually orthogonal directions in the orthogonal subspace of $n_*$,
each from $-\infty$ to $\infty$. We do this next.

The unit normal $n_*$ on the boundary between regions $\cRm$ and $\cR_\tm$
is along the gradient of the negative LLF difference, $-[\ln P(x\mid m)-\ln P(x\mid  \tm)]$,
evaluated at the boundary point $x_*$,
\ba
\label{e30a}
n_*&= -{\nabla_*\ln P(x_*\mid m)-\nabla_*\ln P(x_*\mid \tm)\over \norm{\nabla_* \ln P(x_*\mid m)
-\nabla_*\ln P(x_*\mid m)}}\nonumber\\
 &= {\Sigma_m^{-1}(x_*-x_m)-\Sigma_\tm^{-1}(x_*-x_\tm)\over
 \norm{\Sigma_m^{-1}(x_*-x_m)-\Sigma_\tm^{-1}(x_*-x_\tm)}},
\end{align}
where Eq.~(\ref{e20}) was used to reach the second inequality.
From Eq.~(\ref{e22}) and the fact that $\lambda\approx 1/2$ from expression (\ref{e28})
under conditions used to derive that expression, we see that expression (\ref{e30a}) for
$n_*$ simplifies approximately to the form
\ba
\label{e30}
n_* &= {\Sigma_m^{-1}(x_*-x_m)\over
 \norm{\Sigma_m^{-1}(x_*-x_m)}}\nonumber\\
 &= {\Sigma_m^{-1}(x_*-x_m)\over
 [(x_*^T-x_m^T)\Sigma_m^{-2}(x_*-x_m)]^{1/2}},
\end{align}
where we used the definition of the norm, $\norm{v}^2=v^Tv$, to arrive at the second equality.
We now write $v=v_\parallel+v_\perp$, where
\ba
\label{e31}
v_\parallel &= n_*n_*^T v\nonumber\\
    &=n_*n_*^T\Sigma_m^{-1/2}(x-x_*)+n_*n_*^T\Sigma_m^{-1/2}(x_*-x_m)\nonumber\\
    &=u_m+U_m;\nonumber\\
v_\perp&=v-v_\parallel;
\end{align}
are the projections of $v$ along $n_*$ and in its orthogonal complement, respectively,
with the former subdivided further into two parts, $u_m$ and $U_m$,
\be
\label{e32}
u_m=n_*n_*^T\Sigma_m^{-1/2}(x-x_*); \ \ U_m=n_*n_*^T\Sigma_m^{-1/2}(x_*-x_m),
\ee
in which $U_m$ defines the shift vector between the mean data point $x_m$ within $\cRm$ to the origin, at $x_*$, of the new
coordinate system whose axes are individually scaled by the diagonal elements of the diagonal matrix $\Sigma_m^{-1/2}$.

In view of the definition (\ref{e29}) and the decomposition $v=v_\parallel+v_\perp$, as given in Eq.~(\ref{e31}),
expression (\ref{e20}) may now be exponentiated to arrive at a simplified form for $P(x\mid m)$,
\be
\label{e33}
P(x\mid m)={1\over [(2\pi)^N{\rm det}\Sigma_m]^{1/2}}\exp\left[-(1/2)(v_\parallel^Tv_\parallel+v_\perp^Tv_\perp)\right].
\ee
The integral of $P(x\mid m)$ over $x$ in the decision region $\cR_\tm$,
as we indicated earlier, can now be performed approximately as the product of the integral over the variable $\norm{v_\parallel}$ from
$\norm{U_m}$ to $\infty$ and $(N-1)$ integrals over the remaining $(N-1)$ orthogonal components of $v$,
each of the latter integrals having its limits extended to $\pm\infty$.
The scaling of the coordinate axes in going from the $x$ space to the $v$ space, according to definition (\ref{e29}),
exactly cancels out the determinantal factor in the denominator of expression (\ref{e33}), while the
$(N-1)$ infinite integrals over the orthogonal complement of $n_*$ produce merely an overall factor $(2\pi)^{(N-1)/2}$,
leaving just a single Gaussian integral over $v_\parallel$ to be done. In other words, in the asymptotic limit
the following approximate value may be obtained for the overall multiple integral:
\be
\label{e34}
\int_{\cR_\tm} P(x\mid m)\, dx={1\over (2\pi)^{1/2}}\int_{\norm{U_m}}^\infty\exp[-(1/2)\norm{v_\parallel}^2]\  d\norm{v_\parallel}.
\ee
This result is readily expressed in terms of the complementary error function, as done in the previous section
for the case of purely additive Gaussian data, which leads to the following asymptotically valid
result for the MPE (\ref{e2a}):
\be
\label{e35}
\MPE={1\over 2}\sum_m p_m\, {\rm erfc}\, (\norm{U_m}/\sqrt{2}).
\ee

Numerically a somewhat more accurate form of the asymptotic expression is provided by including
two terms, rather than one, for each value of $m$ in the sum (\ref{e35}), the second term corresponding to
the next nearest decision region at whose boundary $P(x\mid m)$ takes its next highest
boundary value. We employ such an improved approximation for all our numerical results presented in Sec.~6.

\paragraph{Detailed Expression for $\norm{U_m}$}

According to Eq.~(\ref{e32}), the quantity $\norm{U_m}$ is simply the inner product $n_*^T \Sigma_m^{-1/2}(x_*-x_m)$. Using
expression (\ref{e30}) for $n_*$, we may write this inner product as
\be
\label{e36}
\norm{U_m}={(x_*^T-x_m^T)\Sigma_m^{-3/2}(x_*-x_m)\over \left[(x_*^T-x_m^T)\Sigma_m^{-2}(x_*-x_m)\right]^{1/2}}
\ee
In view of the relation (\ref{e23}) between $(x_*-x_m)$ and $\delta x_{\tm m}$ and since $\lambda\approx 1/2$
according to result (\ref{e28}), we may express $\norm{U_m}$ in Eq.~(\ref{e36}) as
\be
\label{e37}
\norm{U_m}={1\over 2}{\delta x_{\tm m}^T{\rm diag}\,\left[ {(\sigma^2+x_m)^2
\over (\sigma^2+\bx_m)^2}\right]\Sigma_m^{-3/2}\delta x_{\tm m}
\over
\left\{\delta x_{\tm m}^T{\rm diag}\,\left[ {(\sigma^2+x_m)^2
\over (\sigma^2+\bx_m)^2}\right]\Sigma_m^{-2}\delta x_{\tm m}\right\}^{1/2}},
\ee
where $\bx_m\defeq (1/2)(x_m+x_\tm)$ is the arithmetic mean of the mean data vectors corresponding to
the two decision regions $\cRm$ and $\cR_\tm$.
Since all the matrices that are sandwiched between $\delta x_{\tm m}^T$ and $\delta x_{\tm m}$ in this
expression are diagonal, they commute with one another and multiply together to yield other diagonal matrices.
With the help of definition (\ref{e19}) for the covariance matrix $\Sigma_m$,
we may thus express the above result as a ratio of two single sums over the data pixels,
\be
\label{e38}
\norm{U_m}={1\over 2}{\sum_{i=1}^N{(\sigma^2+x_{mi})^{1/2}
\over (\sigma^2+\bx_{mi})^2}(\delta x_{\tm mi})^2
\over
\left[\sum_{i=1}^N{1 \over (\sigma^2+\bx_{mi})^2}(\delta x_{\tm mi})^2\right]^{1/2}}.
\ee
In the asymptotic limit, the arithmetic mean, $\bx_m$, of the mean data vectors in the two decision regions that
occurs in this expression may be replaced by either mean data vector, say $x_m$,
without incurring significant error. This would simplify expression (\ref{e38}) somewhat.

\paragraph{Extreme Asymptotic Limit}

As the number of data pixels and the sensitivity of the detector grow large, the MPE sum (\ref{e35})
will be dominated by a single term, namely that value of $m$ for which the argument $\norm{U_m}/\sqrt{2}$ of the complementary
error function is the smallest. In this case the behavior of MPE is asymptotically exponential
with a characteristic exponent, $\nu_\infty$, that takes the value
\ba
\label{e39}
\nu_\infty&\defeq-\lim_{N\to\infty}{\ln \MPE\over N}\nonumber\\
   &=\lim {1\over 2N} \min_m \norm{U_m}^2\nonumber\\
   &= {1\over 8}\min_m \lim_{N\to\infty}{\left[{1\over N}\sum_{i=1}^N{(\sigma^2+x_{mi})^{1/2}
\over (\sigma^2+\bx_{mi})^2}(\delta x_{\tm mi})^2\right]^2
\over
\left[{1\over N}\sum_{i=1}^N{1 \over (\sigma^2+\bx_{mi})^2}(\delta x_{\tm mi})^2\right]},
\end{align}
in which all additive terms in $\ln \MPE$ that scale sub-linearly with $N$ tend to 0 in the limit.
In the pure Gaussian limit of additive noise alone, our results, including those derived earlier in the
section, agree with the corresponding results derived in the previous section, as seen by setting terms
like $\sigma^2+x_m$ to just $\sigma^2$ in all our expressions of the present section. In particular, the exponent
$\nu_\infty$ reduces to the value
\be
\label{e40}
\nu_\infty = {1\over 8\sigma^2}\min_m \lim_{N\to\infty}\left[{1\over N}\sum_{i=1}^N(\delta x_{\tm mi})^2\right]
\ee
in this purely Gaussian noise limit.

\section{Exponentially Peaked Conditional Data Statistics}

The preceding asymptotic analysis of the MPE may be easily extended to any data statistics that may be expressed
naturally in the exponential form,
\be
\label{e41}
P(x\mid m)=\exp[L_m(x; N)],
\ee
where $L_m(x;N)$ is the log-likelihood function (LLF) for the $m$th outcome.
The exponential family of conditional data PDF \cite{Roussas97} is an example of such distributions.
We shall assume the property of exponential peaking for the PDF in the sense
that the LLF is approximately an extensive variable in the asymptotic limit, $N\to\infty$, {\it i.e.}
$\lim_{N\to\infty} L_m(x;N)/N$ is finite.
This assumption must break down in a practical setting since as more and more pixels of
data are included around and outward from the maximum-intensity pixel in the localized image PSF, the less signal per pixel
is expected to be present on average. Depending directly on the extent of the spatial footprint of
the PSF, this must imply, in general, an optimum number of data pixels
beyond which the MPE is expected to show rapidly diminishing improvement and thus essentially to saturate with $N$.
This saturation of the MPE with growing $N$ is evidently present
in expressions like (\ref{e17}), (\ref{e39}), and (\ref{e40})
that tend to saturate with growing $N$.

The boundary of the decision region $\cR_m$ with a neighboring decision region $\cR_{m^\prime}$
under the MAP hypothesis is simply the hypersurface on which $P(x\mid m)\, p_m=P(x\mid{m^\prime})\, p_{m^\prime}$, which in
view of the form (\ref{e41}) of the PDF is equivalent to the equation
\be
\label{e42}
L_m(x;N)+\ln p_m=L_{m^\prime}(x;N)+\ln p_{m^\prime}.
\ee
We now determine that point $x_*$ on the boundary (\ref{e42}) at which $P(x_*\mid m)\, p_m$, or
equivalently $L_m(x;N)+\ln p_m$, is maximized. This is given by a constrained maximization of
this quantity, subject to $x_*$ being on the boundary (\ref{e42}). This amounts, via a Lagrange
multiplier $\lambda$, to maximizing the auxiliary function,
$$ L_m(x;N)-\lambda[L_m(x;N)-L_{m^\prime}(x;N)],$$
which yields the following vanishing-gradient condition:
\be
\label{e43}
(1-\lambda)\nabla_* L_m(x_*;N)+\lambda\nabla_* L_{m^\prime}(x_*;N)=0.
\ee

This is a vector relation from which one can, in principle, evaluate
the $N$ components of the maximizer $x_*=x_*(\lambda;m,m^\prime)$.
The multiplier $\lambda$ must be evaluated self-consistently, however, by substituting the so obtained ``solution" for
$x=x_*$ back into the boundary equation (\ref{e42}). This procedure can be implemented
via an iterative algorithm in which one first starts with an initial guess for $\lambda$, say $\lambda^{(0)}$,
then uses a numerical solver to solve Eq.~(\ref{e43}) for $x_*$, then substitutes the result
in place of $x$ back into Eq.~(\ref{e42}), solving it for an improved value of $\lambda$, say $\lambda^{(1)}$.
This process is repeated as the alternate evaluations of $x_*$ and $\lambda$ are refined from
one iteration to the next. If $x_*^{(n)}$ and $\lambda^{(n)}$ are their values at the $n$th iteration,
one expects them to converge to the correct solution, {\it i.e.}, $\lambda^{(n)}\to\lambda$, $x_*^{(n)}\to x_*$,
as $n\to\infty$.

Depending on the form of the LLFs,
more than one solution $x_*$ is possible. If we label the multiple solutions
by the subscript $s$, running from $1$ to $S$ for $S$ distinct solutions,
then asymptotically the integral of $P(x\mid m)$ over the neighboring decision region $\cR_{m^\prime}$
will, in general, be dominated exponentially by
\be
\label{e44}
\Pi({m^\prime}\mid m)=\max_{s=1,\ldots,S} P(x_{*s}\mid m).
\ee
Under very general conditions and in the extreme asymptotic limit, then, the MPE, as given by the sum (\ref{e5}),
is dominated exponentially by a single term in that sum, that for which the expression (\ref{e44})
is maximized for all possible pairs $(m,m^\prime)$, namely by
\be
\label{e45}
Q_N=\max_{m,{m^\prime},m\neq {m^\prime}}\max_{s=1,\ldots,S} p_m P(x_{*s}\mid m).
\ee
This expression is expected to be equal logarithmically to the MPE in the asymptotic limit of infinitely many pixels,
\be
\label{e46}
-\lim_{N\to\infty}{1\over N}\ln\MPE = -\lim_{N\to\infty}{1\over N}\ln Q_N,
\ee
which via the exponential form of the PDF (\ref{e41}) is equivalent to the asymptotic limit
\be
\label{e47}
-\lim_{N\to\infty}{1\over N}\ln\MPE = -\lim_{N\to\infty}{L_{\tilde m}(\tilde x_*;N)\over N},
\ee
a limit that by our very assumption of extensivity of the LLF is well defined. The quantities
$\tilde x$ and $\tm$ refer to the values of the boundary point $x_*$ and the hypothesis label
for which the double maximization in Eq.~(\ref{e45}) is attained.

We shall now apply our MPE analysis to treat the important task of 3D super-localization
in single-molecule imaging.
Specifically, we shall characterize the MPE incurred in performing this task
using a rotating-PSF imager \cite{SP13} and compare its performance with the conventional clear-aperture imager
for which the PSF in the plane of best focus is of the Airy disk variety.
Since our preceding analysis is completely general, and makes no reference
to a specific application or algorithm, even whether imaging based or not, we are in a position to compare the best
achievable performance, from the Bayesian MPE perspective, of different protocols to achieve such localization.

\section{3D Point-Source Localization Using Conventional and Rotating-PSF Imagers}

The image of a point source produced by an imager in the absence of any sensor or photon noise is simply its
point-spread function (PSF), which we denote as $H(\vec s)$, $\vec s$ being the position
vector in the 2D sensor plane. The PSF must be discretized to reflect image-plane pixelation,
requiring integration of image intensity over the area of each pixel during the recording time to generate what is the mean count of that pixel.
In the presence of photon-number fluctuations, typically described as a Poisson random process,
the count at a pixel fluctuates correspondingly. When detected by a sensor array like a CCD sensor, however,
the read-out process adds further noise, known as additive read-out noise, that may be well approximated, after calibration, as zero-mean Gaussian
noise with a fixed noise variance $\sigma_d^2$.
The recorded count at a pixel then exhibits combined Poisson-Gaussian fluctuations from the two sources of noise,
which, as we noted in Sec.~3, may be described accurately by a pseudo-Gaussian PDF (PGP) with a mean equal to the mean count
at that pixel and a variance equal to the sum of the read-out noise variance and the mean pixel count.
The presence of any background fluctuations, {\it e.g.,} those arising from uncontrolled, largely randomly excited fluorophores
in the imaging volume that add to the fluorescence of the specific molecule of interest in a bioimaging problem,
is yet another source of noise that may be easily accounted for by adding a mean background count, $\bar b$,
to the mean signal count of the PGP at a pixel
and including a corresponding Poisson-noise variance, $\bar b$, in the overall variance. Assuming that the sum total of fluctuations at
the different sensor pixels are statistically uncorrelated, given a specific hypothesis $m$,
the PGP that describes this overall random process is then given by expression (\ref{e18})
with the following values for the mean and variance:
\be
\label{e48}
x_m=s_m+\bar b,\ \ \Sigma_m={\rm diag}(\sigma_d^2+s_m+\bar b),
\ee
where $s_m$ is the vector\footnote{We remap the 2D array of $N\times N$ data pixels into a 1D vector of $N^2$
elements, e.g., by stacking its columns consecutively one atop the next. This enables the use of matrix algebra methods and
validates the results obtained using them in the previous sections.} of conditional mean values of the data pixels, given the hypothesis $m$.
For simplicity, we assume the mean background count to be spatially uniform, so $\bar b$ is simply proportional
to a vector of all 1's.

A rotating-PSF imager is a specific imaging protocol in which a superposition of pure angular-momentum states
of light can be effected through a Fresnel-zone partitioning of the imaging pupil \cite{SP13}. This yields a PSF that
rotates uniformly, in a nearly
shape- and size-invariant manner, with defocus away from its Gaussian image plane. The amount of rotation of the PSF
then encodes the depth $(z)$ information of a point source and its transverse position the $(xy)$ coordinates of the
point source. The rotating PSF, while somewhat more extended spatially in the transverse plane than the conventional clear-aperture
diffraction-limited Airy-pattern PSF, does not, unlike the latter, spread and lose sensitivity even when the defocus-dependent phase
at the edge of the pupil is many waves in magnitude. This represents a trade-off between transverse and axial (depth)
encoding, which we can study and compare for both kinds of imagers under varying strengths of sensor and photon noise using our
Bayesian MPE based metric of performance.

In the formal developments of this section, we keep the PSF as being general that we may choose, as needed, to be either
the rotating PSF or the conventional PSF of a clear aperture without any engineered pupil phase. The
defocus adds a quadratically varying
radial phase in the pupil of form $ \pi(\delta z /\lambda)(\rho/l)^2$, where $\lambda$ is the mean illumination
wavelength, $\delta z$ the defocus distance from the in-focus object plane a distance $l$ from the imaging
pupil, and $\rho$ is the radial distance in the pupil. The phase at the edge of the pupil of radius $R$, namely
$\zeta\defeq \pi(\delta z /\lambda)(R/l)^2$, defines a defocus phase parameter that
we shall henceforth use instead of the actual defocus distance $\delta z$.
Unlike the rotating PSF, the rapid spatial dispersal of the conventional PSF with increasing defocus
should rapidly reduce its sensitivity to encode the depth coordinate of a point source away from the plane of Gaussian focus,
although its tighter spatial footprint in that plane should endow it with a greater sensitivity and resolution
to encode the transverse position of the source, at least in that plane.
It is this fundamental trade-off between the decreased source-localization sensitivity and increased depth of field
for the rotating PSF and their reversal for the conventional PSF that we expect
to capture with the Bayesian MPE analysis. An alternative, minimum mean-squared error (MMSE) based
analysis may also be given to describe this trade-off, but the two Bayesian error analyses are expected to
produce similar conclusions, at least in the highly-sensitive detection limit, as we showed
in Ref.~\cite{SP12}.

Let $I_0H(\vec s-\vec s_m; z_m)$ be the mean intensity in the image of a point source of intensity $I_0$ located at position $(\vec s_m,z_m)$
in the object space. For our discrete representation, we
evaluated our rotating PSF on a finer grid of subpixels than the actual sensor pixel grid, shifted it by the vector amount $\vec s_m$
in the pixel plane, and finally summed over the subpixels constituting each sensor pixel to determine the mean count recorded by the pixel.
We denote such a discrete version of the shifted PSF $H(\vec s-\vec s_m; z_m)$ by $h^{(m)}_{ij}$, with $ij$ being the 2D pixel index,
so the mean count recorded by the $ij$ pixel is $K_0h^{(m)}_{ij}$, where $K_0$ is the intensity $I_0$ expressed in photon count units.

For the case of Gaussian read-out noise alone, we computed the MPE in the form given by Eq.~(\ref{e17})
by noting that the mean count vector $x_m$ is simply the vectorized
version of the mean count array, $K_0h^{(m)}_{ij}$, so the squared norm in that expression is simply the double sum over
a square sensor sub-array of side $\sqrt{N}$ (in pixel units),
\be
\label{e49}
\norm{x_\tm-x_m}^2=K_0^2\sum_{i=1}^{\sqrt{N}}\sum_{j=1}^{\sqrt{N}} \left[ h^{(\tilde m)}_{ij}-h^{(m)}_{ij}\right]^2,
\ee
in which for a given value of $m$, $\tilde m$ labels that source position $m^\prime$ for which $\norm{x_{m^\prime}-x_m}$
is the smallest for all $m^\prime\neq m$. Note that a substitution of expression (\ref{e49}) in the expression (\ref{e17}) for the
MPE immediately exhibits its dependence on the source flux-to-noise ratio (FNR), $K_0/\sigma$. We also note that this asymptotic
expression for the MPE is dominated typically by a single term in the $m$ sum, that for which the norm $\norm{x_{m^\prime}-x_m}$
is the smallest among all possible distinct pairs of source positions, $m\neq m^\prime$, all other terms being exponentially
small in the asymptotic limit of many detected pixels, $N>>1$. As we noted earlier,
this result is quite analogous to the pair-wise minimum Chernoff
distance exponent that characterizes the MPE for M-ary hypothesis testing under asymptotically many IID measurements \cite{LJ97}.

For the more general case of combined Poisson noise of photon fluctuations and Gaussian noise of the CCD read-out process,
the asymptotic form of the MPE is given by a numerically improved modification of expression (\ref{e35}),
as described in the text following that expression. To determine the decision
region $\cR_\tm$ ``closest" to the decision region $\cR_m$ for a given value of $m$, we
required that $U_m$ be the smallest of all $(M-1)$ quantities of the same form as the RHS of (\ref{e38}) in which
$\tm$ is replaced by $m^\prime$ and all $m^\prime\neq m$ are considered. The contribution from the next
nearest decision region was also included for each value of $m$, as needed for the numerically improved
version of the asymptotic expression (\ref{e35}). This procedure was
easily implemented in our Matlab computer code. In our calculations, we
fixed the sensor read-out noise variance at $\sigma_s^2=1$, and the mean background level, taken to be spatially uniform,
was allowed to float with the mean signal strength at the brightest pixel in the conventional in-focus Airk-disk image,
the ratio of the two fixed at 0.1. For such a
uniform mean background level, pixels increasingly farther from the brightest pixel, even for the conventional in-focus image,
see signal levels that get progressively weaker, with the background eventually dominating the signal.
This situation only gets worse when the rotating PSF with its somewhat larger footprint but far better
depth sensitivity is employed for localization. For this case, even with the most compact rotating PSF,
the mean background level was roughly 0.55 of the signal level at the brightest image pixel.
For these reasons, it is sensible to limit attention to only a small sub-image around centered at the brightest pixel,
which we chose to be of area $12\times12$ square pixels.

We divide our 3D localization error analysis into two parts. In the first part,
we calculate the MPE and its asymptotics for 2D transverse super-localization by factors
2x, 4x, 8x, and 16x at a range of focal depths, corresponding to the
defocus phase ranging from 0 to 16 radians in steps of 2 radians, and the reference depth
resolution set at a nominal 1 radian in that phase.
The second part, by contrast, computes the MPE for achieving depth localization enhancements of 2x and 4x, corresponding to
1/2 and 1/4 radian of defocus phase, respectively, at two different defocus phases, 0 and 16 radians, for
the same 4 transverse super-localization factors, 2x, 4x, 8x, and 16x.
These localization error probabilities were computed for the pseudo-Gaussian statistics describing the combined fluctuations
of photon number, CCD read-out, and background illumination. The nominal 3D
localization cell requiring no detailed image data processing defines, for our purposes, the volume over
which the prior may be defined to be uniformly random.
Its two transverse dimensions were chosen to be 4 pixels$\times$ 4 pixels, while its depth dimension, as
we have stated before, was taken to be 1 radian. These somewhat arbitrary choices can be shown to be
consistent with the diffraction limited imaging criterion with respect to the parameters chosen for our PSFs.
Note that much as in actual microscope observations of biological samples, all our images
are thus assumed to be oversampled by the sensor array.

\paragraph{Transverse Localization in 2D}

In our studies, we varied the FNR over eight different values, namely 100, 500, 1000, 2000, 3000, 4000, 5000, and 10000,
which incidentally, since $\sigma_s$ is held fixed at 1, are also the values over which the source photon flux is varied.
The error criterion we employed for reliable localization was that the MPE be less than 5\%,
corresponding to a statistical confidence limit (CL) for a match of better than 95\%. Both exact and asymptotic numerical
evaluations of the MPE, namely of relations (\ref{e1}) and (\ref{e35}), were performed, the former
via a Monte-Carlo sampling approach in which $N_s$ data-vector samples $\{x^{(1)},\ldots,x^{(N_s)}\}$, with $N_s$ sufficiently large,
were drawn according to the
pseudo-Gaussian PDF (\ref{e18}) that approximates $P(x\mid m)$, with the mean and variance given by relation (\ref{e48})
for each source location index $m$. The numerically exact MPE evaluation is predicated on the acceptance
or rejection of each data sample $x$ by testing whether $P(x\mid m)$ is larger than all other $P(x\mid m^\prime)$,
$m^\prime\neq m$, or not. Subtracting from 1 the acceptance rate averaged over all $M$ values of the
hypothesis label $m$ then yields the exact MPE acoording to Eq.~(\ref{e1}). The numerically exact results
we present here were obtained for $N_s=5000$, but our evaluations changed by no more than a percent
even when we increased $N_s$ to 20000 and still larger values in certain cases, giving us excellent
confidence in the accuracy of the numerical results presented here.

\begin{figure}
\centering
\includegraphics[width=3.5in]{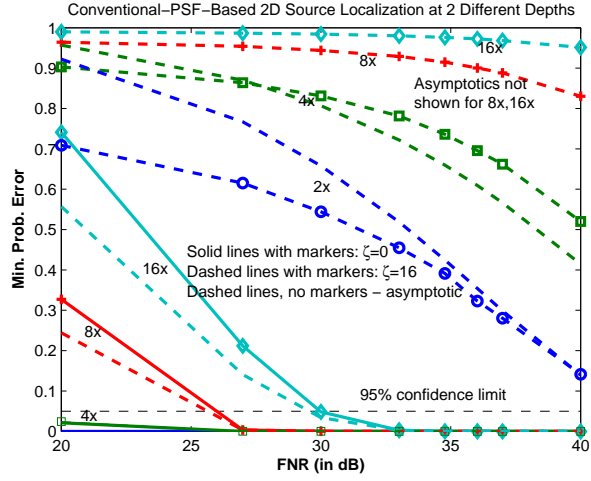}
\caption{\label{fig:f1} Plots of MPE vs. FNR for the conventional imager for two different values of the defocus phase, $\zeta$,
namely 0 and 16 radians. The plots for the various transverse super-localization factors are indicated explicitly both
by means of their different colors, marker symbols, and labels, with the curves in the bottom half referring to $\zeta=0$ (in-focus) and
those in the top half to $\zeta=16$. The corresponding asymptotic values of the MPE, presented by means of dashed line segments,
tend to become more accurate for the smaller values of the MPE. The values of the MPE for 2x super-localization
in the in-focus case are too small to be visible in this figure, while the asymptotic results for the
8x and 16x super-localizations for $\zeta=16$ are too inaccurate to be of any value and are therefore suppressed here.}
\end{figure}

\begin{figure}
\centering
\subfloat[]
{\includegraphics[width=3in]{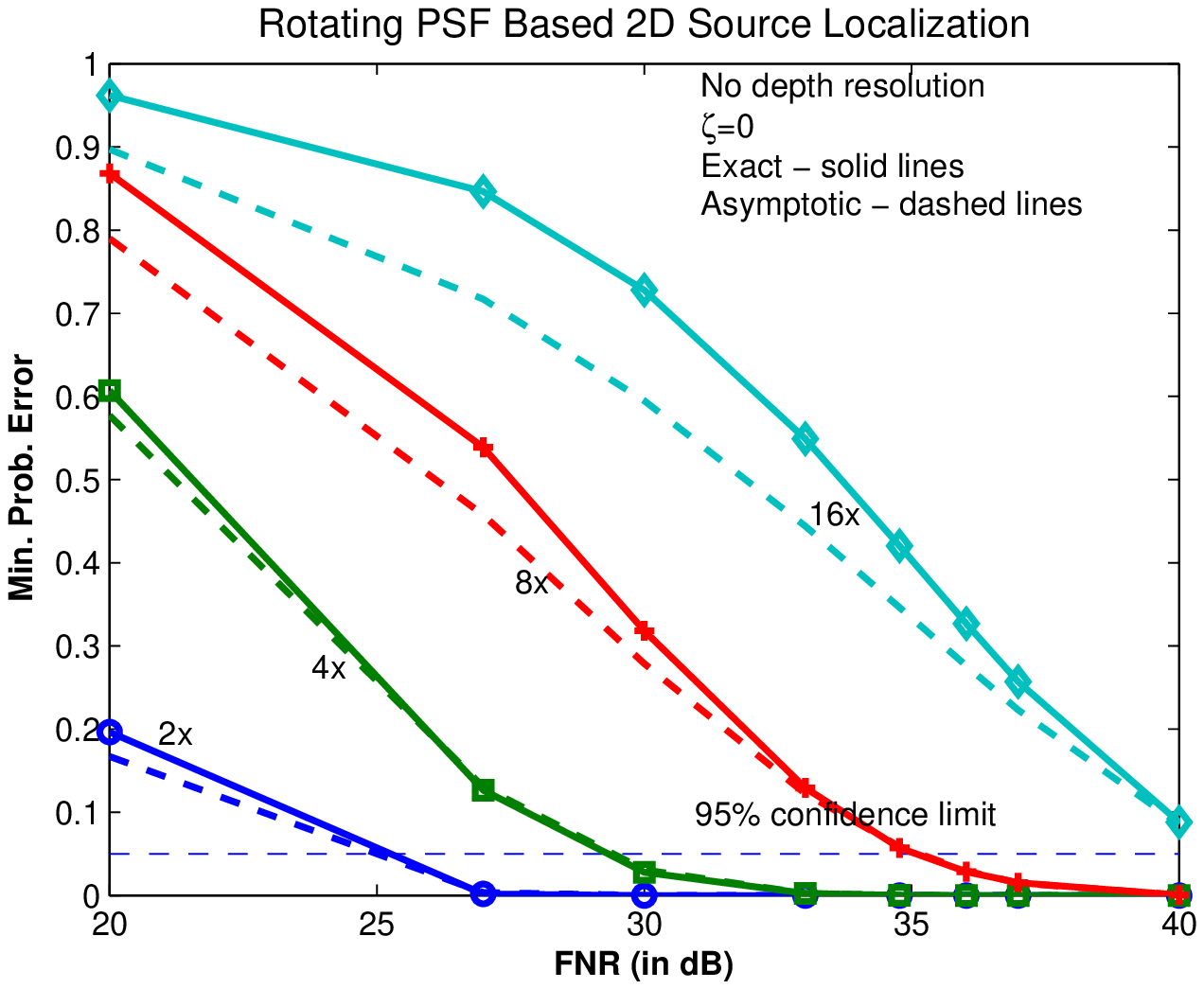}}
\subfloat[]
{\includegraphics[width=3in]{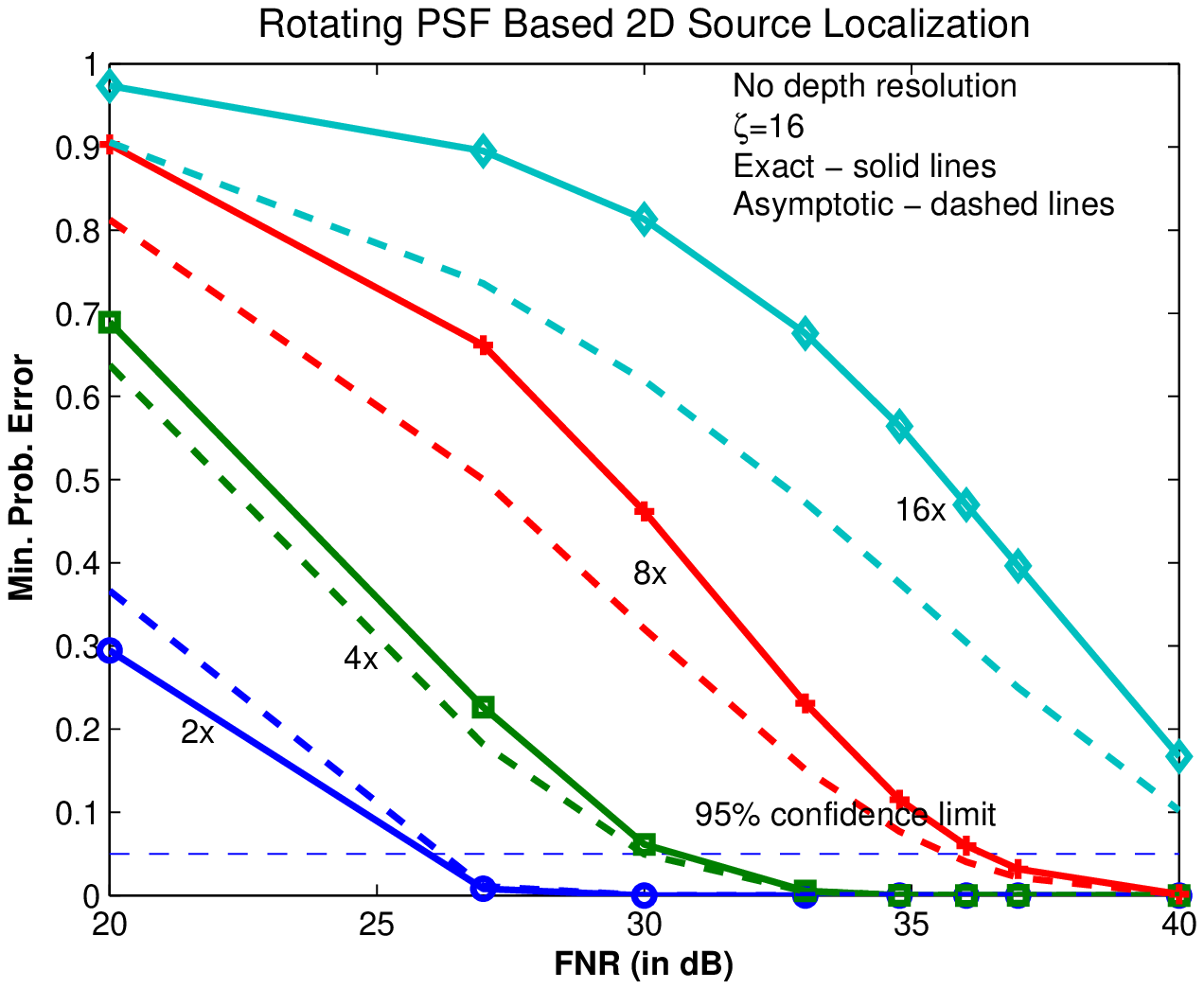}}
\caption{\label{fig:f2} Plots of MPE vs. FNR for the rotating-PSF imager for the same two values of the defocus phase, $\zeta$,
(a) 0 and (b) 16 radians, as in Fig.~1. The plots of the MPE for the various transverse super-localization factors are indicated explicitly both
by means of their different colors, marker symbols, and labels.
The corresponding asymptotic values are presented by means of dashed line segments.}
\end{figure}

We present in Fig.~1 both the exact and asymptotic values of the MPE computed numerically from expressions (\ref{e1}) and
(\ref{e35}) for a number of source FNRs and
for varying degrees of 2D sub-diffractive localization for the conventional imager.
No depth localization was sought in the results displayed in Fig.~1,
but transverse localization at two different depths, namely $\zeta=0$ and $\zeta=16$ radians, was considered.
Since the conventional imager loses sensitivity quickly with increasing depth, the two depth
values are attended by a large disparity in the corresponding values of the MPE. By contrast, the rotating-PSF imager has rather similar
values of the MPE over a considerable depth range, necessitating two different figures, Figs. 2 (a) and 2 (b),
to display them properly for the same two values of the depth.
For the in-focus case ($\zeta=0$), it is clear that the conventional PSF based imager yields significantly smaller
error probabilities than the rotating PSF based imager for the transverse localization task, regardless of
the value of the enhancement factor that we considered here, namely 2x, 4x, 8x, and 16x. This is consistent with the fact that
the conventional PSF has a more compact form than the rotating PSF when in focus. However, the behavior
reverses dramatically at the larger defocus phase, $\zeta=16$ radians, since the rotating PSF maintains its shape
and compactness over such defocus values while the conventional PSF spreads dramatically.
Indeed, in accordance with the approximate depth invariance of the shape and size of the rotating PSF
with increasing defocus, the MPE curves for the rotating-PSF imager are quite comparable at both defocus values.

Based on the comparisons presented in Figs.~1 and 2, it is clear that with increasing defocus
the rotating PSF based imager must outstrip the 2D localization performance of the conventional imager
at some fairly low value of the defocus phase, $\zeta$, but in a manner that depends on the specific values
of FNR and localization enhancement factors chosen. In Figs.~3(a)-(c), we present the comparison of
the MPE based performance of the two imagers as a function of the defocus phase
for a low and a high value of the FNR.
The reversal of the behavior of the error probabilities for 2D super-localization with increasing defocus is
easily seen in the plots of the MPE vs. the defocus phase in these figures. The cross-over of the error curves
for the two kinds of imagers occurs at fairly low defocus values, providing evidence for the fact that the
conventional PSF blurs rapidly with defocus and thus must fail to provide transverse super-localization at all but the
lowest values of the defocus with any reliability. For both imagers, the error probabilities rise, as expected, with the degree of
2D super-localization sought, requiring increasingly larger FNR to keep the MPE acceptably low, but
the rotating-PSF based imager continues to provide excellent localization enhancement at fairly modest values of
the FNR.
\begin{figure}
\centering
\subfloat[]
{\includegraphics[width=3in]{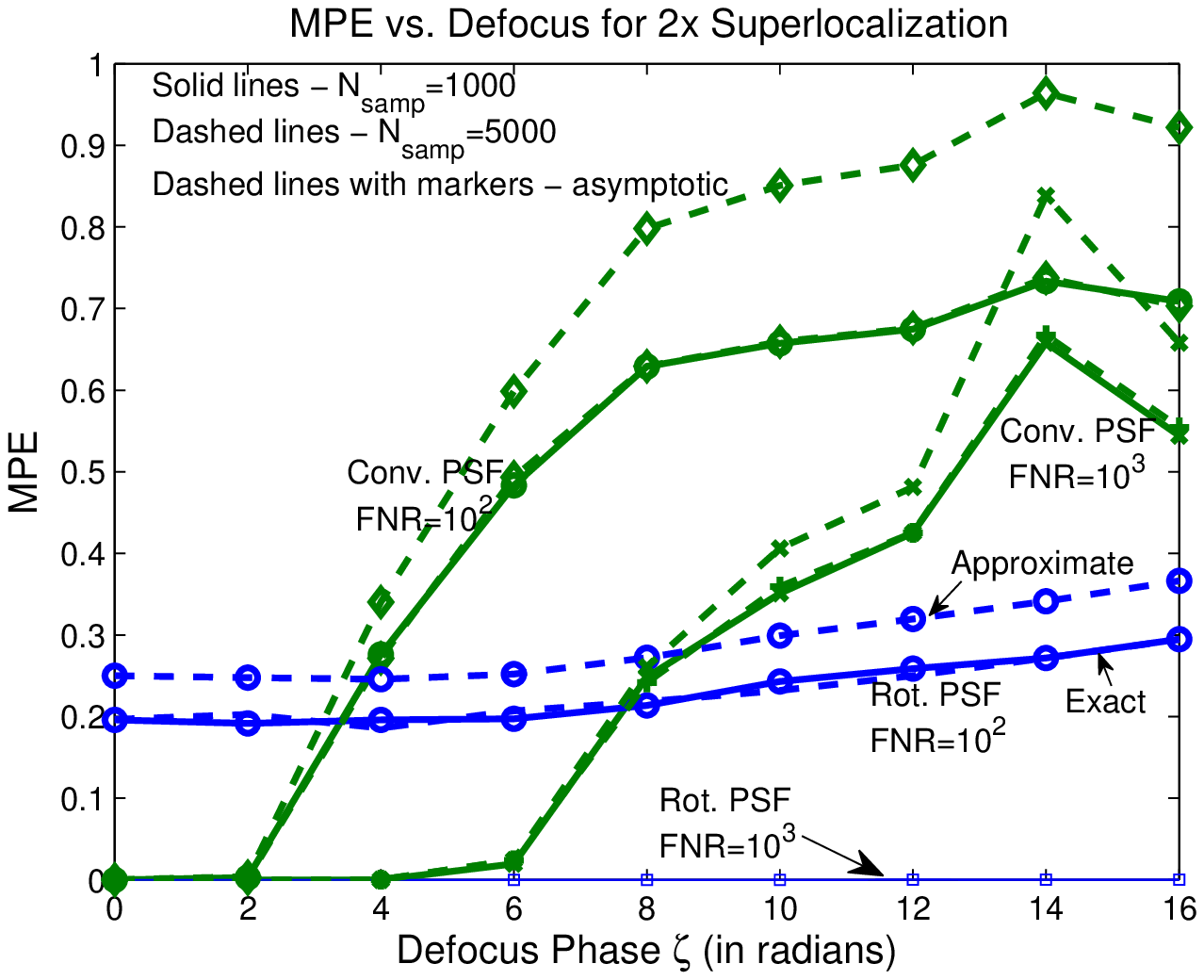}}
\subfloat[]
{\includegraphics[width=3in]{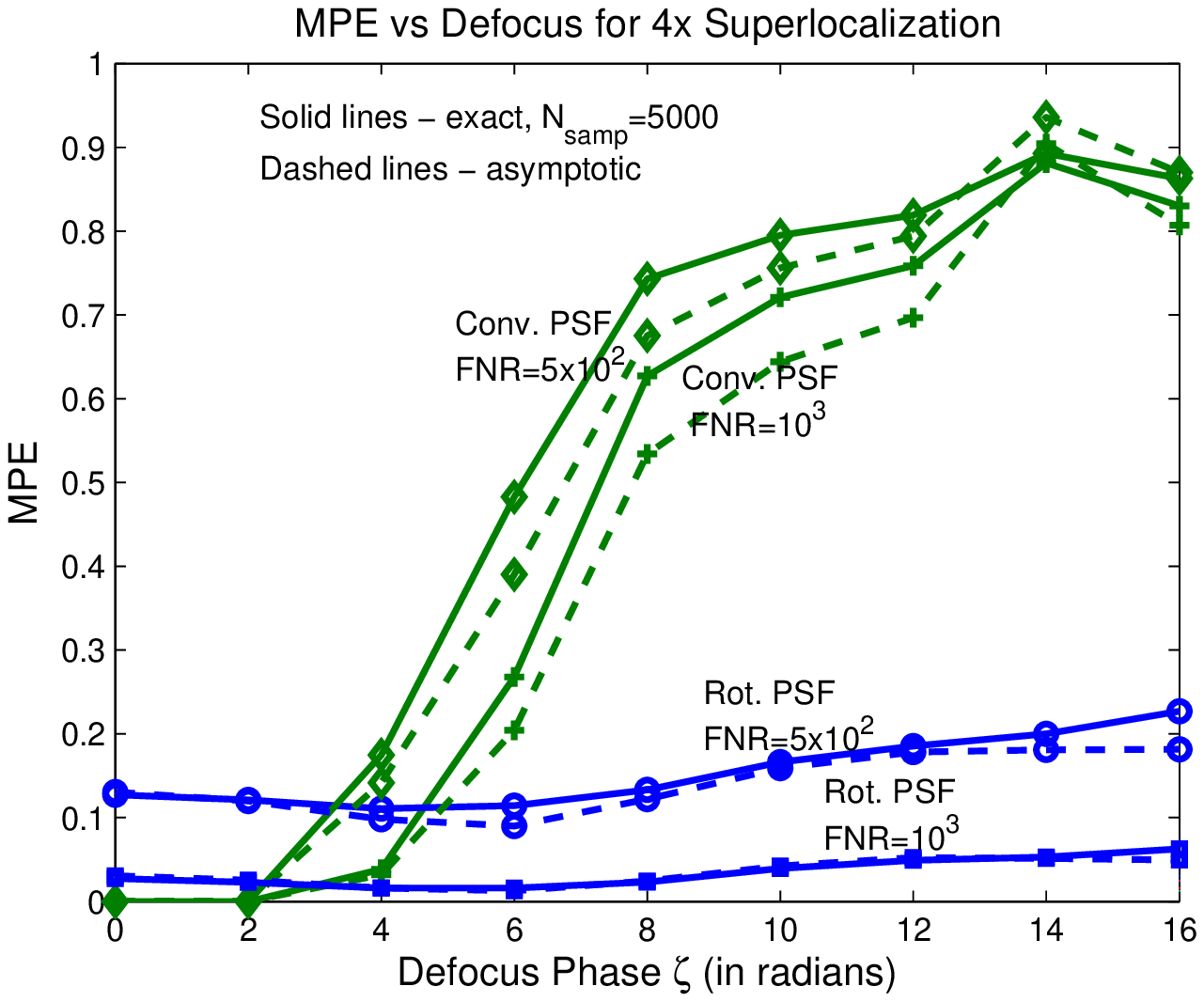}}
\hspace{0cm}
\subfloat[]
{\includegraphics[width=3in]{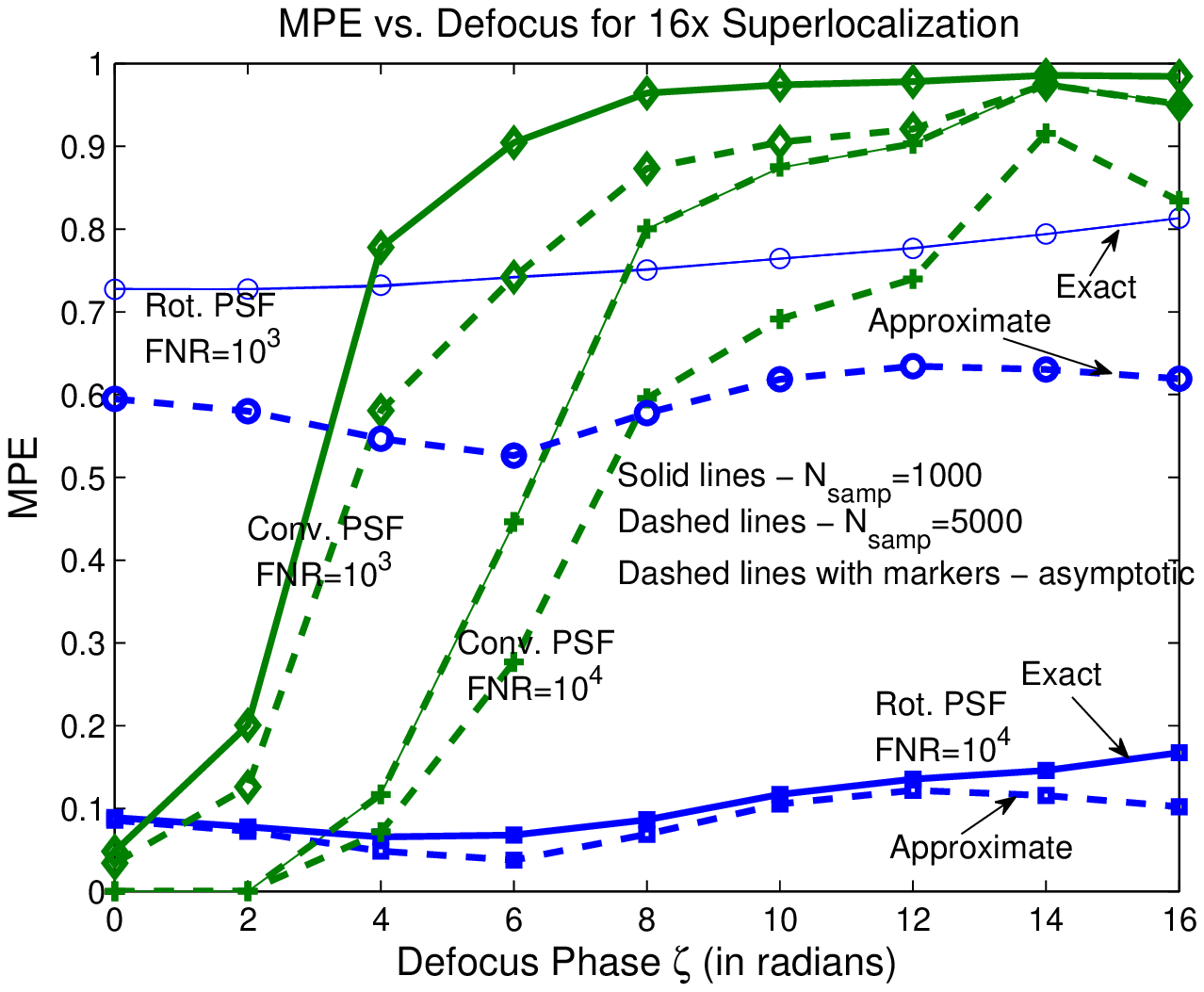}}
\caption{\label{fig:f3} Plots of MPE vs. the defocus phase, $\zeta$, for the rotating-PSF imager (blue curves) and
the conventional imager (green curves) for two different values of FNR. The three subfigures refer to the values (a) 2x; (b) 4x; and
(c) 16x of the 2D super-localization factor. In Figs. 3(a) and 3(c), the closely spaced solid and dashed curves, the latter
barely discernible from the former, refer to two different
values of the number of Monte-Carlo samples drawn to calculate the exact MPE expression (\ref{e1}), namely 1000 and 5000.
The dashed lines with marker symbols display the corresponding asymptotic results.}
\end{figure}

From our graphs of the MPE in Figs.~1 and 2 and other similar graphs, not shown here, for other values of the
defocus phase $\zeta$, we can also read off the
minimum requirements on the source photon number to achieve, at the 95\% statistical confidence limit, or 5\% MPE,
a specific 2D super-localization factor $M_\perp$. We plot in Fig.~4 the minimum source photon number, $K_{min}$,
as a function of $M_\perp^2$ for both the conventional and rotating-PSF imager for four different values of
the defocus phase, namely 0, 4, 8, and 16 radians. As expected, for the rotating-PSF imager, $K_{min}$, for each value of $M_\perp$,
increases rather modestly as it is defocused more and more over the full 16-radian range,
while for the conventional imager defocused localization even for $\zeta=4$ requires roughly
double the $K_{min}$ needed for the former imager operating at $\zeta=16$.
Only at best focus, $\zeta=0$, does the conventional imager deliver a better localization performance
at more modest photon numbers. Its PSF spreading is
simply too dramatic, as it is defocused from its best focus,
for it to stay competitive with the rotating-PSF imager.
Although our results that we display here were obtained under the conditions of the mean background to
the peak brightness of the conventional in-focus image ratio at 10\% and negligible sensor noise,
we have verified this comparative performance of the two imagers more generally.
\begin{figure}
\centering
\includegraphics[width=3.5in]{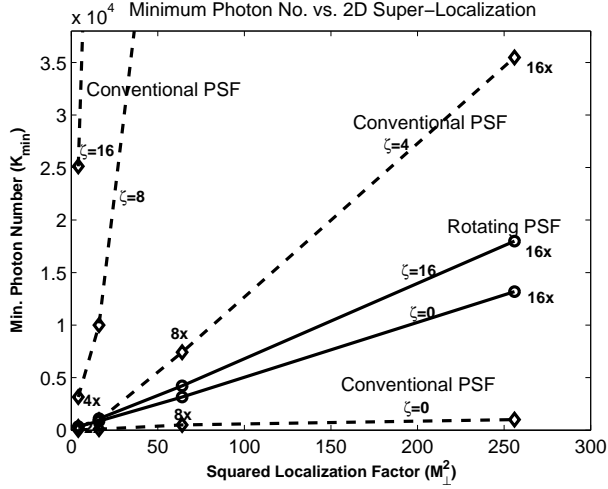}
\caption{\label{fig:f4} Plots of $K_{min}$ vs. $M_\perp^2$ for the two imagers for two different values of the defocus phase, $\zeta$,
namely 0 and 16 radians. The dashed lines are based on extrapolations of our numerical results from those shown in the previous figures.}
\end{figure}

The approximately linear form of the plots in Fig.~4 for both imagers
confirms the approximately quadratic dependence of the
minimum source strength on the degree of super-localization sought at any value of the defocus.
This conclusion is quite consistent with previous mean-squared-error (MSE) based analyses \cite{TLW02,ORW04}
in the signal dominated regime of operation.

\paragraph{Full 3D Localization}

We now address the problem of localizing the spatial position of a point source in all three dimensions
by means of an imager that employs either a rotating PSF or the conventional PSF. As is well appreciated \cite{RWO05}, the
axial, or depth, localization provided by the conventional imager is rather poor for a source at
the plane of best focus because of a lack of first-order sensitivity of the PSF relative to the defocus at this plane.
The rotating PSF imager, however, has no such insensitivity, and is helped further by its
ability, in sharp contrast with the conventional imager, to maintain a tight PSF profile over a large defocus range.
We examine, via our present MPE analysis, these attributes of the rotating-PSF imager to achieve full 3D localization.

In Fig.~5, we display, for the rotating PSF imager, the MPE as a function of the source photon number for the same 10\%
background level used in the previous figures, but now for different degrees of transverse
and depth localizations at two different depths, $\zeta=0$ [Fig.~5(a)] and $\zeta=16$ [Fig.~5(b)].
As expected, the MPE increases with increasing demand on the degree of depth localization
from 2x to 4x for each of the transverse localization enhancement factors, 2x, 4x, 8x, and 16x,
at each of the two reference depths. The rather modest differences between the overall
behaviors presented in the two figures are a result of the approximate invariance
of the rotating PSF shape and size across a large range of defocus. By contrast, the next two figures, Figs.~6(a)
and 6(b), which present the corresponding results for the conventional imager, indicate
a much larger error profile, even at zero defocus for the reason we have stated earlier, namely
the first-order insensitivity of such imager in providing any depth discrimination at
the in-focus plane. However, since the Bayesian error-probability analysis provides a
global sensitivity metric, the higher order sensitivity of conventional imagers
to yield depth discrimination is fully accounted for in our results.
\begin{figure}
\centering
\subfloat[]
{\includegraphics[width=3in]{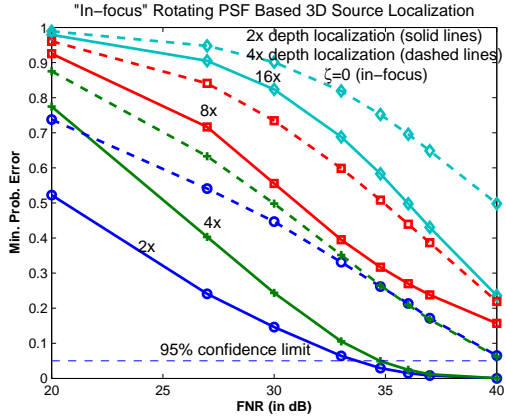}}
\subfloat[]
{\includegraphics[width=3in]{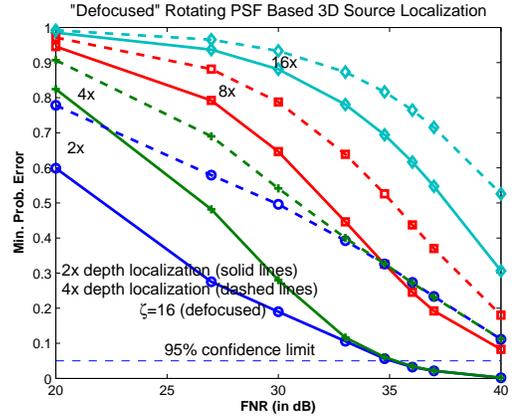}}
\caption{\label{fig:f5} Plots of MPE vs. source signal strength (in dB) for two different values of the defocus phase, (a) $\zeta=0$
and (b) $\zeta=16$, for two different axial and four different transverse localization enhancement factors
for the rotating-PSF imager. The latter are indicated by different marker symbols, namely circle for 2x, + for 4x, square for 8x, and
diamond for 16x, while the former are indicated by the line type, solid for 2x and dashed for 4x.}
\end{figure}

\begin{figure}
\centering
\subfloat[]
{\includegraphics[width=3in]{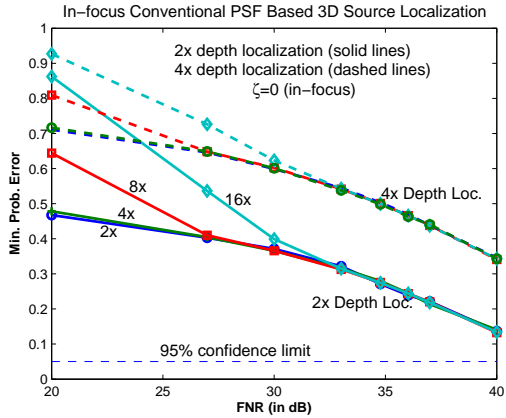}}
\subfloat[]
{\includegraphics[width=3in]{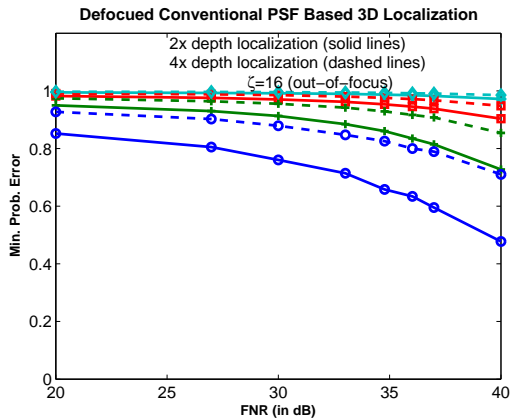}}
\caption{\label{fig:f6} Same as in Figs. 5(a) and 5(b), except for the conventional imager.}
\end{figure}

The competition between axial and transverse localizations is also evident in these figures.
With an increasing signal strength, the behavior of the MPE is strongly influenced by any changes
in the requirement of transverse localization, rising as it does with increasing values of the latter,
but ultimately at sufficiently high strengths the MPE is limited by the requirement of depth localization alone,
as seen in the asymptotic behavior of the various MPE curves at a fixed depth localization
but for different transverse localization factors. This behavior is quite universal for both
imagers at sufficiently large signal strengths for each depth localization,
but it is particularly noticeable for the in-focus ($\zeta=0$) conventional imager
for which the 2x and 4x transverse super-localization curves are essentially indistinguishable
at the higher (4x) depth super-localization demanded over the full range of signal strengths considered here.

\section{Concluding Remarks}

The present paper has considered an asymptotic analysis of
the Bayesian problem of multi-hypothesis testing, designed specifically
to treat the fidelity of point-source localization with sub-diffractive error
that is based on image data. We apply our exact and approximate
analyses of the minimum probability of error (MPE) in discriminating among
$M_\perp^2\times M_\parallel$ possible outcomes of the source position inside
an elementary base resolution volume that was subdivided uniformly into $M_\perp^2\times M_\parallel$
subvolumes. The transverse and axial localization
enhancement factors, $M_\perp$ and $M_\parallel$, were chosen to have values 2,4,8,16 and 2,4, respectively,
The image data were drawn from a small sub-image, here $12\times 12$ square pixels,
centered around the brightest pixel in the full image. Two different imaging systems, one based
on conventional clear-aperture imaging and the other on a phase engineered rotating PSF imaging, were
compared for their MPE-based potential to achieve 3D source super-localization
The MPE was calculated for a number of different signal strengths of the source, ranging from 100 to $10^6$ photons,
at a mean background level that was pegged at 10\% relative to the brightest pixel in the conventional in-focus image.

In the signal-dominated regime for which we have presented our detailed calculations here,
we confirmed a number of conclusions drawn by previous researchers using mean-squared error (MSE) based analyses
about the minimum source signal strength needed for localizing a point source spatially with precision
exceeding what is nominally possible, namely of order $\lambda^2/NA^2\times \lambda/NA^2$ which we regard as
our base resolution volume. In particular, we showed a quadratic ($M_\perp^2$) dependence of the minimum source
strength needed to achieve a transverse localization improvement factor
of $M_\perp $ at a statistical confidence limit of 95\% or better. The agreement in the predictions of MSE and MPE based analyses
in the high-sensitivity asymptotic limit is a consequence of the equivalence of the two metrics in that
limit \cite{SP12}.

From our calculations of the MPE for combined 3D transverse-axial localization, we
demonstrated an interplay between axial and transverse localization
enhancements. We found that the sought axial localization enhancement typically serves to limit the 3D localization
at sufficiently high signal strengths, at which all of the transverse localization improvements from 2x to 16x
entail no added error cost.
The reduced sensitivity of the conventional imager to perform any depth resolution at zero defocus,
noted previously in a local error-bound analysis \cite{RWO05},
is also borne out well in our global MPE based Bayesian analysis.

It is worth noting that the MPE
is the best error performance one can achieve from the view point of error probability in a Bayesian detection protocol,
and most actual localization algorithms are likely to be sub-optimal from this perspective. Reasons for this sub-optimality
are many, not the least of which are both an incomplete identification of statistical sources of error and
their imperfect statistical characterization. Effort should be devoted primarily in mitigating
these systematic sources of error before employing the MAP estimate for localization. Without such mitigation,
the performance bound presented by our MPE analysis may seem overly optimistic under realistic FNR conditions.

\section{Acknowledgments}

Helpful conversations with R. Kumar, S. Narravula, J. Antolin, Z. Yu, H.
Pang, K. Lidke, R. Heintzmann, and R. Ober are gratefully acknowledged. The work reported here was supported in part by AFOSR
under grant numbers FA9550-09-1-0495 and FA9550-11-1-0194.

\end{document}